\documentclass[aps,preprint,preprintnumbers,nofootinbib,showpacs,prd]{revtex4-1}
\usepackage{graphicx,color,amsmath,amsfonts,multirow}
\usepackage{epsfig}
\usepackage{rotating}
\usepackage{array}
\usepackage{graphicx}
\usepackage{multirow}
\usepackage{amssymb}
\usepackage{latexsym}

\newcommand{\beq}     {\begin{equation}}
\newcommand{\eeq}     {\end{equation}}
\newcommand{\bea}     {\begin{eqnarray}}
\newcommand{\eea}     {\end{eqnarray}}
\newcommand{\bit}     {\begin{itemize}}
\newcommand{\eit}     {\end{itemize}}
\newcommand{\ben}     {\begin{enumerate}}
\newcommand{\een}     {\end{enumerate}}
\newcommand{\bad}{\begin{array}{ccc}}
\newcommand{\ea}{\end{array}}

\newcommand{\bbt}{\bibitem}
\newcommand{\lsim}{\mathrel{\mathop{\kern 0pt \rlap
  {\raise.2ex\hbox{$<$}}}
  \lower.9ex\hbox{\kern-.190em $\sim$}}}
\newcommand{\gsim}{\mathrel{\mathop{\kern 0pt \rlap
  {\raise.2ex\hbox{$>$}}}
  \lower.9ex\hbox{\kern-.190em $\sim$}}}
  
\newcommand{\no}     {\nonumber}

\newcommand{\br}{{\rm B}}

\newcommand{\tb}      {{\tan\beta}}

\newcommand{\cb}      {\cos\beta}
\renewcommand{\sb}      {\sin\beta}
\newcommand{\ca}      {\cos\alpha}
\newcommand{\sa}      {\sin\alpha}

\newcommand{\gm}      {\gamma}
\newcommand{\Gm}      {\Gamma}
\newcommand{\sg}      {\sigma}

\newcommand{\bb}      {{b \bar{b}}}

\newcommand{\llp}      {{\ell^+\ell^-}}
\newcommand{\rr}      {{\gamma\gamma}}
\newcommand{\ttau}      {{\tau\tau}}
\newcommand{\hsm}      {{h_{\rm SM}}}
\newcommand{\ehgg}      {{\epsilon_{gg\rm F}}}
\newcommand{\esvbf}      {{\epsilon_{\rm VBF}}}
\newcommand{\chtot}{{C^{ h}_{\rm tot}}}

\newcommand{\fig}{\begin{figure}}
\newcommand{\ef}{\end{figure}}

\newcommand{\ggf}{{gg {\rm F}}}
\newcommand{\vbf}{ {\rm VBF}}
\newcommand{\gev}{{\;{\rm GeV}}}
\newcommand{\tev}{{\;{\rm TeV}}}

\def\lgeff{{\mathcal L}_{\rm eff}}
\def\lgeffa{{\mathcal L}_{\rm eff}^A}

\newcommand{\wtR}{{\widetilde{R}\,}}

\newcommand{\ahf}{{\cal A}^{h}_{1/2}}
\newcommand{\ahv}{{\cal A}^{h}_{1}} 
\newcommand{\aaf}{{\cal A}^{A}_{1/2}}

\newcommand{\catot}{{C^{A}_{\rm tot}}}

\newcommand{\yh}      { {\widehat{y}} }

\begin{document}
\title{ 
Comprehensive study of two Higgs doublet model \\
in light of the new boson with mass around 125 GeV 
}

\author{Sanghyeon Chang$^a$, Sin Kyu Kang$^b$, Jong-Phil Lee$^b$,
\\ Kang Young Lee$^c$,
Seong Chan Park$^d$, Jeonghyeon Song$^a$}
%\email{jhsong@konkuk.ac.kr}

\affiliation{
$^a$Division of Quantum Phases \& Devices, School of Physics, 
Konkuk University, Seoul 143-701, Korea\\
$^b$School of Liberal Arts, Seoul-Tech, Seoul 139-743, Korea\\
$^c$Department of Physics Education, 
Gyeongsang National University,
Jinju 660-701,
Korea\\
$^d$Department of Physics, Sungkyunkwan University,
Suwon 440-746, Korea
}

\date{\today}
\begin{abstract}
The recent discovery of a new boson of mass roughly 125 GeV
has been reported by the ATLAS and CMS collaborations.
Although its signals in various decay modes resemble those of the 
standard model (SM) Higgs boson,
observed are the combinations of entangled information about the production,
decay rates and total decay width of the new boson.
In addition, some decay channels show
non-negligible deviation from the SM expectation, such as
the $2\sigma$ excess in
the diphoton channel.
In the four types (I, II, X and Y) of two Higgs doublet models,
we perform the global $\chi^2$ fit in three scenarios:
(i) the new boson is the light CP-even Higgs boson $h^0$;
(ii) it is the heavy CP-even Higgs boson $H^0$;
(iii) the signals are from degenerate $h^0$ and the pseudoscalar $A^0$.
Considering other phenomenological constraints such as flavor physics,
electroweak precision data, and the LEP search for the Higgs boson,
we find that the the first
scenarios in Type II and Type Y models actually provide better or
similarly good fit to the data than the SM.
All the other models are excluded at 95\% C.L..
\end{abstract}

\maketitle
\section{Introduction}
\label{sec:introduction}

%\key{Higgs has been discovered with mass 125 GeV}{ Kang}
Recently, the CMS and ATLAS collaborations announced the discovery of 
a new boson of mass around 125 GeV 
in their search for the Higgs boson~\cite{ATLAS-Higgs,CMS-Higgs,ATLAS-7,CMS-7}.
The analysis of several different Higgs decay channels
shows that the properties of this new particle
are consistent with those of the elementary Higgs boson of the standard model (SM).
This observation has independent support from the Tevatron, 
where the excess of the observed data over background 
can be interpreted as a signal of the Higgs boson of mass ranging
between 115~GeV and 135~GeV~\cite{tevatron}.

%\key{Summary of 11 discovery channels for the SM Higgs boson}{Park}
However, the issue of whether it is indeed the SM Higgs boson is still open.
The Higgs boson is produced via five main channels at the LHC:
%There are five main mechanisms for Higgs production in pp collisions considered by CMS and %ATLAS:
the gluon-gluon fusion ($gg$F), the vector boson fusion (VBF), 
the associated vector boson production ($Wh, Zh$), and 
the associated production with top quarks ($t\bar{t} h$)~\cite{handbook11,handbook12}.
The largest production cross section of the Higgs boson in the SM
is from the gluon fusion.
The second largest cross section is from the VBF process, which 
is about 8\% (5.5\%) of the gluon fusion one 
without kinematic cuts applied at 7 TeV (8 TeV)~\cite{handbook11,handbook12}.
The production mechanisms are inferred from the event selection,
which are tagged by dijet for VBF, by leptons for $Wh$ and $Zh$, 
or untagged for the gluon fusion.

Various decay channels of the Higgs boson open
as the Higgs boson mass is the sublime of 125 GeV. 
Measurable are $b\bar{b}$, $\tau\tau$, $WW^*$, $ZZ^*$, and $\rr$ decay modes.
In particular, the loop-induced couplings with $\rr$ and $gg$, 
which allow one of the highest sensitivities,
have almost maximum branching ratio for the Higgs mass around 125 GeV.
The precision measurement of these branching ratios shall open a new indirect channel
for probing new physics.
In fact, the CMS and ATLAS collaborations probe the five most sensitive 
decay modes (i.e. $\gamma\gamma, ZZ, WW,\tau\tau, bb$) 
and other less sensitive sub-channels~\cite{CMS-7, ATLAS-7}. 

Albeit too early to draw decisive conclusions,
the current data on the Higgs boson signals 
show some deviations from the SM expectation.
One of the most significant deviations 
is the enhancement in the $h\rightarrow \gamma \gamma$ rate for both the gluon fusion and VBF production.
Both ATLAS and CMS experiments have also seen some deficits in the $b\bar{b}$ and $\ttau$
channels.
Their implications on various new physics models have been studied very extensively in the 
literature~\cite{np,fingerprinting}.
We note the reader that this new physics study should come with caution.
There still exist theoretical QCD uncertainties~\cite{Godbole},
especially in the gluon fusion production cross section $\sg_\ggf$
despite the dedicated work on the calculation at the NLO~\cite{ggf:nlo}
and at NNLO~\cite{ggf:nnlo}.

From a theoretical perspective,
the modification of Higgs sector  has long been expected 
in order to resolve the gauge hierarchy problem.
Many new physics models 
have extended Higgs sector.
One of the simplest extensions is the two-Higgs-doublet-model~\cite{2hdm1,2hdm2,2hdm3}. 
In order to suppress CP-violation in the Higgs sector as well as
tree-level flavor-changing-neutral-current (FCNC),
we consider CP-conserving 2HDM
with an additional discrete symmetry such that one fermion couples with only one Higgs doublet~\cite{Glashow:Weinberg}.
In 2HDM, there are five physical scalar bosons:
two neutral CP-even scalar, one CP-odd scalar, and two charged scalar bosons  
($h^0, H^0, A^0, H^\pm$). 
There are four types of 2HDM satisfying these conditions,
Type I, Type II, Type X and Type Y models~\cite{2hdm:type1,2hdm:type2,2hdm:type3}.
The collider phenomenology of Type I and II models
are well established in previous studies~\cite{2hdm1,2hdm2,2hdm3,Kominis:1994fa,Aoki:2009ha,Branco}. 
In light of new Higgs data from the LHC, mostly Type II model has been re-examined
in the parameter space of 
$\tb\geq 1$~\cite{Giardino:2012dp,Chang:2012tb,Carmi:2012yp,Carmi:2012in,Cheon:2012rh,Gori}.

In this paper, we perform the global fit to the Higgs signal strength data
comprehensively in all four types of 2HDM.
In addition to the ordinary approach where the observed new boson
is the light CP-even neutral Higgs boson (Scenario-1),
we consider two more scenarios, motivated by the electroweak precision data.
In Scenario-2,
the observed scalar is the heavy Higgs scalar ($H^0$) 
and the light scalar $h^0$ is not observed at the LEP
because its production cross section is small enough~\cite{LEP,LEP2}.
Scenario-3 assumes that the observed signals are 
from almost degenerate state of the light scalar ($h^0$) and the pseudoscalar ($A^0$)~\cite{Elvira}.
The question whether any model is better in explaining the Higgs signal than the SM Higgs boson
is to be answered by globally fitting.
Other phenomenological constraints from the LEP bounds, 
the $\rho$ parameter, and flavor physics
are also considered in the analysis.

The paper is organized as follows.
In Sec.~\ref{sec:review},
we give a brief review of 2HDM. 
The effective Lagrangian describing the Higgs couplings to fermions are to be summarized 
for four types of 2HDM.
In Sec.~\ref{sec:data}, we define and summarize the current Higgs signal rates,
and present the effective Lagrangian and parameters.
Section \ref{sec:results} deals with the results of the global $\chi^2$ fit 
in four type of 2HDM with three scenarios. 
Concluding remarks follow in Sec.~\ref{sec:conclusions}.

\section{Brief review of 2HDM}
\label{sec:review}
A two-Higgs-Doublet-Model (2HDM) is one of the minimal extensions of the SM Higgs sector
where a single Higgs doublet provides mass for the up-type and down-type fermions.
This economical setup is relaxed to allow two complex doublets of the Higgs fields:
\bea
 H_u = \left( \begin{array}{c}
             H_u^+ \\
             \dfrac{v_u+H_u^0+iA_u^0}{\sqrt{2}} \end{array} \right )\;, \qquad
 H_d = \left( \begin{array}{c}
             H_d^+ \\
            \dfrac{v_d+H_d^0+iA_d^0}{\sqrt{2}}   \end{array} \right )\;,
\eea
where $v_u$ and $v_d$ are non-zero vacuum expectation value (VEV),
which defines $\tan\beta=v_u/v_d$.
The electroweak VEV of the SM 
is related via $v=\sqrt{v_u^2+v_d^2}$.

In a 2HDM, there are five physical scalars,
the light CP-even scalar $h^0$,
the heavy CP-even scalar $H^0$, the CP-odd scalar $A^0$,
and two charged Higgs bosons $H^\pm$.
Neutral Higgs bosons are
\bea
h^0 &=& \sqrt{2}(H_d^0 \sin \alpha -H_u^0 \cos\alpha),  \\\nonumber
H^0 &=& -\sqrt{2}(H_d^0 \cos \alpha +H_u^0 \sin \alpha), \\\nonumber
A^0 &=&\sqrt{2}( A_d^0 \sin \beta -A_u^0 \cos \beta) .
\eea
Since the SM Higgs boson is 
\bea
\hsm
&=& h^0\sin(\alpha -\beta) -H^0\cos(\alpha-\beta),
\eea
the $h^0$ becomes identical with $\hsm$
if $\sin(\alpha -\beta) = 1$.
This is called the decoupling limit.

\begin{table}
\caption{\label{tab:Yukawa}The normalized Yukawa couplings of 
the up-type quark $u$, the down-type quark $d$, and the charged lepton $\ell$,
with neutral Higgs bosons.}
\begin{ruledtabular}
%\begin{tabular}{l|lll| lll|lll}
\begin{tabular}{l|ccc| ccc|ccc}
& $\yh_u^{h}$ &$\yh_d^{h}$ &$\yh_\ell^{h}$ &
   $\yh_u^{H}$ &$\yh_d^{H}$ &$\yh_\ell^{H}$ &
   $\yh_u^{A}$ &$\yh_d^{A}$ &$\yh_\ell^{A}$ \\
\hline
Type I & $\frac{\ca}{\sb}$ &$\frac{\ca}{\sb}$ &$\frac{\ca}{\sb}$ &
   $\frac{\sa}{\sb}$ &$\frac{\sa}{\sb}$ &$\frac{\sa}{\sb}$ &
   $\cot\beta$ &$-\cot\beta$ &$-\cot\beta$ \\
   \hline
Type II & $\frac{\ca}{\sb}$ &$-\frac{\sa}{\cb}$ &$-\frac{\sa}{\cb}$ &
   $\frac{\sa}{\sb}$ &$\frac{\ca}{\cb}$ &$\frac{\ca}{\cb}$ &
   $\cot\beta$ &$\tan\beta$ &$\tan\beta$ \\
      \hline
Type X & $\frac{\ca}{\sb}$ &$\frac{\ca}{\sb}$ &$-\frac{\sa}{\cb}$ &
   $\frac{\sa}{\sb}$ &$\frac{\sa}{\sb}$ &$\frac{\ca}{\cb}$ &
   $\cot\beta$ &$-\cot\beta$ &$\tan\beta$ \\
         \hline
Type Y & $\frac{\ca}{\sb}$ &$-\frac{\sa}{\cb}$ &$\frac{\ca}{\sb}$ &
   $\frac{\sa}{\sb}$ &$\frac{\ca}{\cb}$ & $\frac{\sa}{\sb}$ &
   $\cot\beta$ &$\tan\beta$ & $-\cot\beta$\\
\end{tabular}
\end{ruledtabular}
\end{table}

Naive extension of the SM into 2HDM yields large contributions
to FCNC
since two Yukawa matrices
from two Higgs doublets cannot be simultaneously diagonalized in general.
One effective way to suppress FCNC at the leading order
is to impose a discrete symmetry such that one fermion couples with only one Higgs doublet~\cite{Glashow:Weinberg}.
According to the charges of the quarks and leptons under the discrete symmetry,
there are four types of 2HDM: 
Type I, Type II, Type X, and 
Type Y models~\cite{2hdm:type1}.
We parameterize the Yukawa interactions with $h^0$, $H^0$, and $A^0$ as
\bea
 {\cal L}_{\rm Yuk} =
 - \sum_{f=u,d,\ell} \frac{m_f}{v}
 \left(
 \yh_f^h\bar{f}f h^0 
 +
 \yh_f^H\bar{f}f H^0
 - i
 \yh_f^A \bar{f}\gm^5 f A^0 
 \right) ,
\eea
where the effective couplings of $\yh_f^{h,H,A}$ in
four types of 2HDM are summarized in Table \ref{tab:Yukawa}.

In a general 2HDM, there are six phenomenological parameters:
\bea
M_{h^0}, \quad M_{H^0}, \quad M_{A^0}, \quad M_{H^\pm},\quad \alpha, \quad \tb.
\eea
Various observables at low energy put significant constraints on the model parameters.
The first constraint is from the electroweak precision data,
especially from the $\rho$ parameter~\cite{Gunion:1989we}.
The current data is~\cite{pdg}
\bea
\Delta\rho \equiv \rho^{\rm obs}-\rho^{\rm SM}\approx 0.0002 \pm 0.0007.
\eea
New contributions to $\rho$ in 2HDM are  
\bea
\Delta \rho &=& \frac{\sqrt{2} G_F}{(4\pi)^2}\left\{
F_{\Delta \rho}(M_A^2,M_{H^\pm}^2)
-\sin^2(\alpha-\beta)\left[
F_{\Delta \rho}(M_A^2,M_{H}^2)-
F_{\Delta \rho}(M_H^2,M_{H^\pm}^2)
\right] \right. \nonumber \\
&& \left. -\cos^2(\alpha-\beta)\left[
F_{\Delta \rho}(M_h^2,M_A^2)-
F_{\Delta \rho}(M_h^2,M_{H^\pm}^2)
\right]
\right\},
\eea
where
\bea
F_{\Delta \rho}(M_1^2,M_2^2)\equiv
\frac{1}{2}(M_1^2+M_2^2)-\frac{M_1^2M_2^2}{M_1^2-M_2^2}\ln \frac{M_1^2}{M_2^2}.
\eea
One of the simplest ways to suppress $\Delta \rho$ is to assume
almost degenerate masses of $A^0$ and $H^0$.
Another interesting condition for very small 
$\Delta \rho$ is 
$M_{h^0}\simeq M_{A^0}$, $M_{H^0}\simeq M_{H^\pm}$, and 
$\sin^2(\alpha-\beta) \simeq 1$. 

The second constraint on the model parameters
is from the perturbativity of Yukawa couplings of top and $b$ quarks:
$(y_t)^2 \lsim 4\pi$ and $(y_b)^2 \lsim 4 \pi$~\cite{2HDM:perturbation:bound}.
It limits
the value of $\tan \beta$ between 0.29 and 
50~\cite{Aoki:2009ha,Branco,2HDM:perturbation:bound}.
More severe constraints, especially on $\tb$ and $M_{H^\pm}$, are from various flavor physics
such as purely leptonic decays of $B$ and $D$ mesons, $\Delta M_B$, $b\to s \gm$,
and $Z\to \bb$~\cite{flavor}.
Among four types of 2HDM, Type II is most strongly constrained,
while Type I and Type X are least constrained.
Data on $b\to s \gm$ exclude small mass region of charged Higgs boson mass
($M_{H^\pm} \gsim 300\gev$) for Type II and Type Y models,
and small $\tb$ for Type I and Type X models.
$\Delta M_{B_d}$ excludes small $\tb$ for all types of 2HDM.
Only the Type-II model has significant upper bound on $\tb$ from $D_s \to \tau\nu_\tau$,
{\it e.g.}, $\tb\lsim 50 $ for $M_{H^\pm}=600\gev$.

These phenomenological constraints affect the parameter scan.
Considering the strong bounds from flavor physics,
most studies of 2HDM in the literature assume $\tb>1$ and heavy charged Higgs boson.
If 2HDM is not the final theory but an effective way to describe the Higgs sector,
we can relax the constraint on $\tb$~\cite{Gunion:2011}.
A larger theory for new physics may evade the flavor constraint,
{\it e.g.}, through the cancellation of the charged Higgs contributions to various
FCNC.
In this study, we consider two cases, 
{\tt Unconstrained} and {\tt Flavor-constrained} cases.
For the {\tt Unconstrained} case,
we scan all the parameter space of $-\pi/2<\alpha<\pi/2$
and $0.1 < \tb < 50$.
For the {\tt Flavor-constrained} case,
we assume rather heavy charged Higgs boson like $M_{H^\pm} =1\tev$,
which limits $\tb$ as 
\bea
\label{eq:flavor}
\hbox{ {\tt Flavor-constrained}  } && \quad \hbox{Type-I and Type-X: }~~ \tb> 1, \\ \no
 && \quad \hbox{Type-II and Type-Y: }~~ \tb> 0.5.
\eea

In four types of 2HDM, 
we consider the following three scenarios:
\begin{description}
\item[Scenario-1] The observed signal is from the light CP-even neutral Higgs boson $h^0$.
\item[Scenario-2] The new boson is the heavy CP-even $H^0$,
and the light CP-even $h^0$ has been missed.
\item[Scenario-3] The observed signal is from two almost degenerate $h^0$ and $A^0$.
\end{description}
Naturally all three scenarios suppress the contribution to $\Delta\rho$.
The first two scenarios explain the data by a single particle resonance.
We do not consider the scenario where the new boson is the CP-odd scalar boson $A^0$,
since it is highly disfavored by the presence of VBF process.
The third scenario is allowed by EWPD if $\sin^2 (\alpha-\beta) \simeq 1$.
The question is whether this bizarre scenario is allowed by the observed Higgs signal.
We label each by Model $A$-$i$, where $A={\rm I,II,X,Y}$ denotes the 2HDM type,
and $i=1,2,3$ the suggested scenario.

\section{Data on the LHC Higgs search and effective couplings for signals}
\label{sec:data}
\subsection{LHC Higgs signals}
\label{subsec:data}

\begin{table}
\caption{\label{tab:R}Summary of the LHC Higgs signals}
\begin{ruledtabular}
\begin{tabular}{|l|l|l|}
&  ATLAS and CMS  & CMS\\
\hline
$7\tev$ & $\wtR_\rr^\ggf=1.66 \pm 0.50, \quad \wtR_{WW}^\ggf = 0.58 \pm 0.41$
& $\wtR_\rr^\ggf=1.66 \pm 0.50,\quad \wtR_{WW}^{Vh}=2.75 \pm 2.96$\\
& $\wtR_{WW}^\ggf = 0.58 \pm 0.41,\quad \wtR_{ZZ}^\ggf = 0.79 \pm 0.41$
& $\wtR_\ttau^\vbf= -1.61 \pm 1.25,\quad \wtR_{\ttau}^{Vh} =0.659 \pm 3.07$\\
& $\wtR_\ttau^\ggf = 0.75 \pm 1.02,\quad \wtR_{bb}^{Vh} = 0.62 \pm  1.09$ & \\ \hline
$8\tev$ & $\wtR_\rr^\ggf=1.69 \pm 0.44,\quad \wtR_\rr^\vbf = 1.34 \pm 0.94$
& $ \wtR_{WW}^\vbf =1.34 \pm 1.82,\quad \wtR_\ttau^\ggf = 2.14 \pm 1.48$ \\
& $\wtR_{WW}^\ggf = 1.38 \pm 0.49,\quad \wtR_{ZZ}^\ggf = 0.85 \pm 0.40$ &
$\wtR_\ttau^\vbf= -1.73 \pm 1.25,\quad \wtR_{\bb}^{Vh} =0.43 \pm 0.80$\\
\end{tabular}
\end{ruledtabular}
\end{table}

In this subsection,
we parameterize the observed Higgs signal.
Useful parameterization for the observed signal in the Higgs search at the LHC
is the ratio of the observed event rate of a specific channel to the SM expectation,
given by
\bea
R^{\tt production}_{\tt decay}
\equiv
\frac{\sum_{j}\sg (pp\to j \to h) \times \br(h\to {\tt decay})|_{\rm observed}}
{\sum_{j} \sg (pp\to j \to h) \times \br(h\to {\tt decay})|_{\rm SM}},
\eea
where $j$ runs over all Higgs production channels
satisfying a specific ``{\tt production}" event selection,
${\tt production} = {gg \rm F}, {\rm VBF}, Vh$ and
$ {\tt decay} = \rr, WW, ZZ, bb,\ttau$.
As in many studies, we identify $R$'s with the signal strength modifier $\hat\mu = \sigma/\sg_{\rm SM}$
which maximizes the likelihood function of the test statistics.
We denote the observed Higgs rates by $\wtR$'s, and the expected ones by $R$'s.
The 18 Higgs signals on various $\wtR^{\tt production}_{\tt decay}$ 
reported by the ATLAS
and CMS collaborations
at the LHC with $\sqrt{s}=7\tev$
and $8\tev$ are summarized in Table \ref{tab:R}.
When combining the ATLAS and CMS data,
we assume that the signal rate $\wtR$
in a given channel follows a Gaussian
distribution.
The correlations in combinations of different channels and/or experiments 
are to be neglected~\cite{fingerprinting,error:sum}.

The superscript {\tt production} in $R^{\tt production}_{\tt decay} $ could be misleading
especially for the VBF production of the Higgs boson.
Any event is included in this class
if passing the dijet tag designed to select the VBF mainly through two forward jets~\cite{CMS:H:rr}.
Non-negligible numbers of the events from the gluon fusion production
pass the dijet tag since dijets can be radiated through QCD interaction.
The gluon fusion cross section in the SM is
about 13 times larger than the VBF cross section.
The dijet-tagged gluon fusion is about 38\% of the tagged VBF in the SM~\cite{CMS:H:rr}.
Therefore we have 
\bea
R_{ii}^{\rm VBF}
&=&
\frac{\sigma(pp \to hjj) {\rm Br}(h \to i i ) }
{\sigma(pp \to \hsm jj) {\rm Br}(\hsm \to i i ) }
\\ \nonumber
&=&  
\frac{\ehgg \cdot \sigma(gg\to h)+\esvbf \cdot  \sigma(VV\to h) }
{\ehgg \cdot \sigma(gg\to \hsm )+\esvbf \cdot  \sigma(VV\to \hsm) }
\cdot \left|
\frac{{\rm Br}(h \to i i )}{{\rm Br}(\hsm \to i i ) }\right|^2.
\eea
Here $\ehgg$ and $\esvbf$ are the efficiencies 
of the gluon fusion and the VBF, respectively, to pass the VBF selection cuts.
Other production channels, such as the  gluon fusion
and $Vh$ using lepton tag,
are to be considered as a single production channel.

We shall perform the global $\chi^2$ fit of model parameters to the observed Higgs signal strength,
with $\chi^2$ defined by
\bea
\label{eq:chi2:def}
\chi^2 = \sum_{i=1}^N \frac{(R_i - \wtR_i)^2}{\sigma_i^2},
\eea
where $i$ runs for all the Higgs search channels,
and for the error $\sigma_i$ we use the $1\sg$ systematic errors reported 
by the ATLAS and CMS collaborations.

\subsection{Single particle scenarios}
\label{subsec:single}

In the scenarios where the observed signals are from a single particle resonance,
2HDM effects
are parameterized by the effective Lagrangian of \cite{Carmi:2012yp,Carmi:2012in}
\bea
\lgeff &= &
c_V {2 m_W^2 \over v} h  \,  W_\mu^+ W_\mu^- + c_V  {m_Z^2 \over v} h  \, Z_\mu Z_\mu 
\\ \no
&& - c_{b} {m_b \over v } h  \,  \bar b b  - c_{\tau} {m_\tau \over v } h \, \bar \tau \tau
- c_{c} {m_c \over v } h  \, \bar  c c 
- c_{t} {m_t \over v } h  \, \bar  t t
\\ \nonumber &&
+ c_{g} {\alpha_s \over 12 \pi v} h  \, G_{\mu \nu}^a G^{a \mu \nu}
+ {c}_{\gamma } { \alpha \over \pi v} h  \, A_{\mu \nu} A^{\mu \nu}\,,
\eea
where $h=h^0$ or $h=H^0$.
For $m_h =125$ GeV, the SM values are
\begin{eqnarray}
\label{eq:2}
c_{V,\rm SM} = \left. c_{f,\rm SM} \right|_{f=t,b,c,\tau}=1\,,\qquad c_{g,\rm SM} \simeq 1\,, 
\qquad {c}_{\gamma,\rm SM} \simeq -0.81\,.
\end{eqnarray}

Very good approximations for $R^{\tt production}_{\tt decay}$
in terms of the effective couplings are
\bea 
R_{\rr}^{gg \rm F}
&=&
\left|
\frac{c_g {c}_\gm}{{c}_{\gm,\rm SM} \chtot }
\right|^2,
\qquad
R_{ii}^{gg \rm F} 
=
\left|\frac{c_g c_i}{ \chtot }\right|^2, 
\qquad
R_{ii}^{Vh}
=
\left|\frac{c_V c_i }{ \chtot }\right|^2,
\\ \no
R_{\rr}^{\rm VBF}
&=&
\widehat{R}_{\rm VBF}^h
\left|
\frac{{c}_\gm}{{c}_{\gm,\rm SM} \chtot }
\right|^2,
\qquad
R_{ii}^{ \rm VBF}
=
\widehat{R}_{\rm VBF}^h
\left|\frac{c_i}{ \chtot }\right|^2, 
\eea
where $\chtot = \sqrt{\Gm_{\rm tot}^{h}/\Gm_{\rm tot}^{\hsm} }$, 
$i=W,Z,\tau,b$,
and the effective VBF production rate relative to the SM expectation $\widehat{R}^h_{\rm VBF}$ is
\bea
\label{eq:Rhat}
\widehat{R}_{\rm VBF}^h
=\frac{\ehgg\cdot |c_g|^2 \sigma^{SM}_{gg\to h}
+\esvbf\cdot  |c_V|^2 \sigma^{SM}_{\rm VBF} }
{\ehgg \cdot \sigma^{SM}_{gg\to h}  + 
\esvbf \cdot \sigma^{SM}_{\rm VBF}} .
\eea
Note that, as well as the Higgs coupling parameter for the given decay mode,
the total decay width affects $R$'s.

Without additional fermions or charged vector bosons,
$c_g$ and ${c}_\gm$ are determined by $c_{t,b,c,\tau,V}$.
The loop-induced effective couplings of a CP-even scalar with a gluon pair 
and a photon pair are 
\bea
\label{eq:cg}
c_g &=& \sum_{q=t,b,c}c_q \ahf(x_q) ,
\\ \no
{c}_\gm &=& 
\frac{2}{9} \sum_{u=c,t}c_u \ahf(x_u) + \frac{1}{18}c_b \ahf(x_b) + 
\frac{1}{6} c_\tau \ahf(x_\tau) - c_V \ahv (x_W),
\eea
where $x_i = m_h ^2/4 m_i^2$.
The loop functions $A^{h}_{1/2,1}$ are
\bea
\ahf(x) &=& \frac{3}{2x^2} \left [ (x-1)f(x) + x \right ],
\\ 
\no
\ahv(x) &=& \frac{1}{8 x^2}
\left[
3 (2 x-1) f(x) +2 x + 2 x^2
\right],
\eea
where
\begin{eqnarray}
\label{eq:3}
f(x)
= \left\{ \begin{array}{lll}
{\rm arcsin}^2\sqrt{x} && x \le 1 
\\ 
-\frac{1}{4}\left[\log\frac{1+\sqrt{1-x^{-1}}}{1-\sqrt{1-x^{-1}}}-i\pi\right]^2 && x > 1 \end{array}\right.\,.
\end{eqnarray}
The loop-induced $\gm$-$\gm$-$h$ vertex has two main contributions
from the top quark and the $W$ boson.
In the SM, the top quark contribution has opposite sign of the $W$ contribution.
If either of $c_t$ or $c_V$ changes the sign,
the diphoton signal is enhanced. 

\subsection{Degenerate scenario}
\label{subsec:effective:vertex}

We consider the case in which two scalar bosons $h^0$ and $A^0$ 
cooperate to explain the signal of the new boson.
This is possible when $h^0$ and $A^0$ are almost degenerate.
%We do not consider another case where $h$ and $H$ are almost degenerate.
It is worthwhile to notice that the pseudoscalar cannot give rise to the contribution to
the Higgs production via VBF.
In this case, the effective Lagrangian is
\bea
\lgeffa &= &
c_V {2 m_W^2 \over v} h \,  W_\mu^+ W_\mu^- + c_V  {m_Z^2 \over v} h \, Z_\mu Z_\mu 
- c_{b} {m_b \over v } h \,  \bar b b  - c_{\tau} {m_\tau \over v } h \, \bar \tau \tau
- c_{t} {m_t \over v } h \, \bar  t t 
\\ \nonumber &&
+ c_{g} {\alpha_s \over 12 \pi v} h \, G_{\mu \nu}^a G^{a \mu \nu}
+ {c}_{\gamma } { \alpha \over \pi v} h \, A_{\mu \nu} A^{\mu \nu}
\\ \no &&
- a_{b} {m_b \over v } A \,  \bar b \gm_5 b  - a_{\tau} {m_\tau \over v } A \, \bar \tau \gm_5 \tau
- a_{t} {m_t \over v } A \, \bar  t \gm_5 t 
+ a_{g} {\alpha_s \over 12 \pi v} A \, G_{\mu \nu}^a G^{a \mu \nu} 
+ a_{\gamma } { \alpha \over \pi v} A \, A_{\mu \nu} A^{\mu \nu}
\,.
\eea
The pseudoscalar $A^0$ couples with photons and gluons through
\bea
a_g &=& a_t \aaf (x_t) + a_b \aaf(x_b) ,
\\
a_\gm &=& \frac{2}{9} a_t \aaf(x_t) + \frac{1}{18}a_b \aaf(x_b) + 
\frac{1}{6} a_\tau \aaf(x_\tau),
\eea
where
\bea
\aaf(x) &=& \frac{3}{2 x} f(x).
\eea

The relevant Higgs event rates are the same except for the following 4 channels:
\bea
R_{\rr}^{gg \rm F}
&=&
\left|
\frac{c_g {c}_\gm}{{c}_{\gm,\rm SM} \chtot }
\right|^2
+
\left|
\frac{a_g {a}_\gm}{{c}_{\gm,\rm SM} \catot }
\right|^2
,
\\
R_{\ttau}^{gg\rm F}
&=& 
\left|\frac{c_g c_\tau}{ \chtot  }\right|^2
+
\left|\frac{a_g a_\tau}{ \catot  }\right|^2,
\\
R_{\rr}^{\rm VBF}
&=&
\widehat{R}_{\rm VBF}^h \left|{ c_{\gamma} \over  c_{\gamma,SM} \chtot }\right|^2
+
\widehat{R}_{\rm VBF}^A \left|{a_{\gamma} \over  c_{\gamma,SM} \catot }\right|^2,
\\ 
R_{\ttau}^{\rm VBF}
&=&
\widehat{R}_{\rm VBF}^h \left|{ c_{\tau} \over  \chtot }\right|^2
+
\widehat{R}_{\rm VBF}^A \left|{a_{\tau} \over  \catot }\right|^2,
\eea
where $\catot = \sqrt{ \Gm^A_{\rm tot}/\Gm^{h_{\rm SM}}_{\rm tot} }$,
$\widehat{R}_{\rm VBF}^h$ is in Eq.~(\ref{eq:Rhat}),
and $\widehat{R}_{\rm VBF}^A$ is
\bea
\label{eq:Rhat:A}
\widehat{R}_{\rm VBF}^A
=\frac{\ehgg\cdot |a_g|^2 \sigma^{SM}_{gg\to h}
 }
{\ehgg \cdot \sigma^{SM}_{gg\to h}  + 
\esvbf \cdot \sigma^{SM}_{\rm VBF}} .
\eea

\section{Results}
\label{sec:results}

\subsection{Scenario-1}
If the observed new boson is $h^0$, 
the effective couplings are
\bea
\label{eq:h:c}
c_V &=& \sin (\beta-\alpha),
\quad
c_b =\yh_d^{h},\quad
c_\tau = \yh_\ell^{h},
\quad 
c_t =c_c = \yh_u^{h},
\eea
where $\yh_d^{h}$, $\yh_\ell^{h}$,  and $\yh_u^{h}$
in terms of $\alpha$ and $\tb$ for Type I, II, X, and Y are in Table \ref{tab:Yukawa}.

{\renewcommand{\tabcolsep}{25pt}
\begin{table}
\caption{\label{tab:Scenario1}Best-fit points for Scenario-1 
in Type I, II, X, and Y 2HDM.
For the {\tt Unconstrained} parameter space, we take $-\pi/2<\alpha<\pi/2$
and $\tb>0.1$.
For the {\tt Flavor-constrained}, we only allow $\tb> 0.5$ for Type II and Type Y,
and $\tb>1.0$ for Type I and Type X.}
%\begin{ruledtabular}
\begin{tabular}{|l|l|l|}
\hline
 &{\tt Unconstrained} &{\tt Flavor-constrained}\\
 \hline
2HDM Type & $(\chi^2_{\rm min},~\alpha,~\tb)$ & $(\chi^2_{\rm min},~\alpha,~\tb)$ \\ 
\hline \hline
Type I-1 & $(16.15,\, 1.38,\, 0.21)$ & $(30.12,\,  -0.97,\, 1.02)$ \\ \hline
Type II-1 \checkmark & $(16.16,\, 1.21,\, 0.36)$ & $(20.31,\, 0.96,\, 0.50)$ \\ \hline
Type X-1 & $(15.24,\, 1.19,\, 0.27)$ & $(28.55,\, -0.001,\, 49.76)$ \\ \hline
Type Y-1 & $(15.80,\, 1.38,\, 0.21)$ & $(23.89,\, 1.06,\, 0.51)$ \\ \hline
\end{tabular}
\end{table}
}

With the data 
in Table \ref{tab:R}, we perform $\chi^2$ analysis.
The SM Higgs boson has 
\bea
\label{eq:SM:chi2}
\left.
\chi^2_{\rm SM}  
\right|_{{\rm d.o.f.}=18}=23.04.
\eea
In Table \ref{tab:Scenario1},
we present the best-fit point on $\alpha$ and $\tb$,
for the {\tt Unconstrained} case ($\tb\in [0.1,50]$)
and  {\tt Flavor-constrained} case as in Eq.(\ref{eq:flavor}).
In the {\tt Unconstrained} case
all four types of 2HDMs have smaller $\chi^2_{\rm min}$ than the SM,
for small $\tb \sim 0.2-0.3$
and large $\alpha \sim 1.2-1.4$
with $\chi^2_{\rm min}$ around $15-16$.
The LHC Higgs search signal alone prefers small $\tb$ and large $\alpha$.

For the {\tt Flavor-constrained} parameters as in Eq.~(\ref{eq:flavor}),
the $\chi^2_{\rm min}$ values in four types of 2HDM increase.
Note that $\chi^2|_{99\% \;\rm C.L.}=34.8$, $\chi^2|_{95\% \;\rm C.L.}=28.9$
for 18 degrees of freedom.
Type I-1 model has been known as weakly constrained by flavor physics,
allowing quite light charged Higgs boson.
However, this mode is excluded at 95\% C.L. by the LHC Higgs signals, if it should satisfy the
flavor physics.
In addition, Type X-1 model with the FCNC constraints is also almost excluded at 95\% C.L.
Flavor-allowing Type Y-1 model is as good as the SM for the LHC Higgs signal. 
The Type II-1 model, even with flavor constraints,
has smaller $\chi^2_{\rm min}$ value than the SM, 
although the difference is not significant enough to claim that
the LHC Higgs signal definitely prefers Type II-1 model.

%%%%%%%%%%%%%%%%%%%%%%%%%%%%%%%%%%%%%%%%%%%%%%%%%%%%%%%%%%%%
\begin{figure}[h]
\begin{center}
\includegraphics[width=0.49\linewidth]{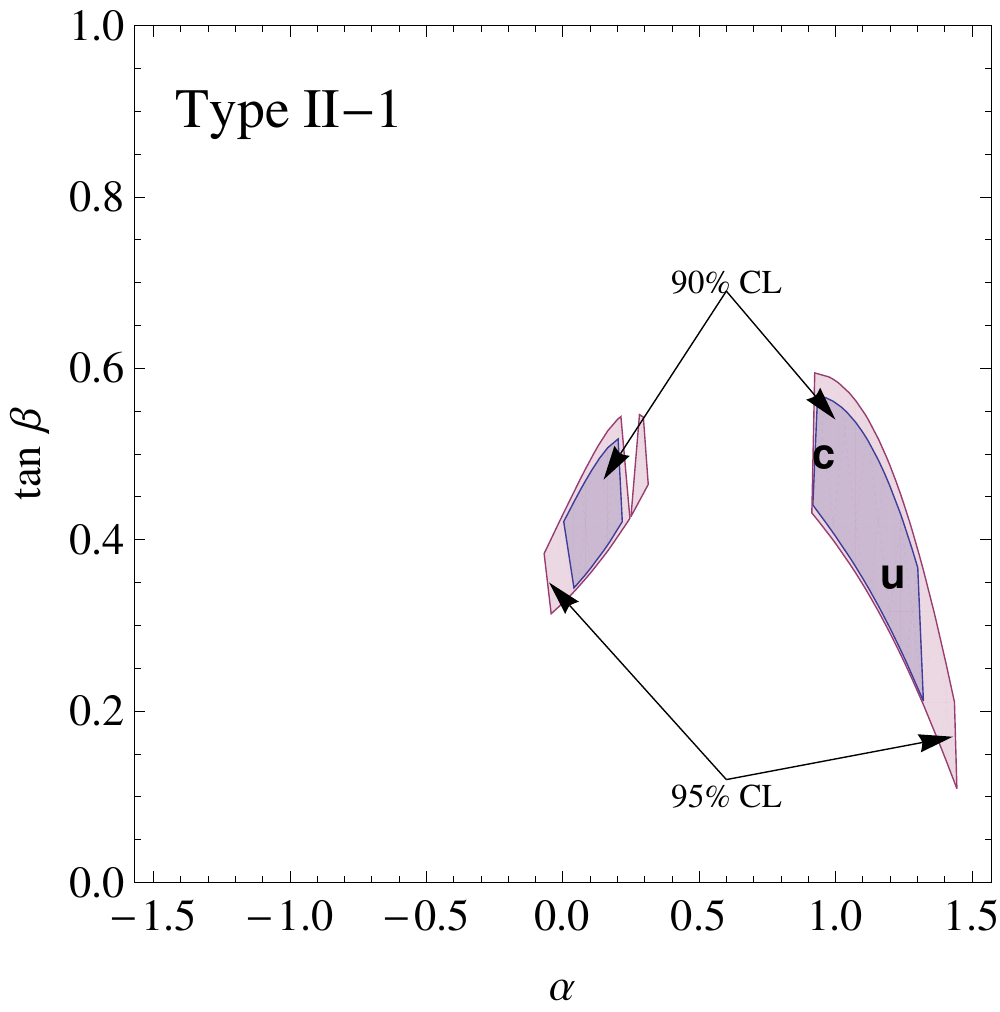}
\end{center}
\caption{Contours for 90 (95)$\%$ in Type II-1 model 
where $h=h^0$.
The best-fit occurs at the point marked by {\tt\bf u} in the {\tt Unconstrained} parameter scan,
and at the point marked by {\tt\bf c} in the {\tt Flavor-constrained} parameter scan.}
\label{fig:t2s1contour}
\end{figure}

The Type II-1 model is the best scenario.
Figure \ref{fig:t2s1contour} shows the contours 
for the allowed regions of parameter space $(\alpha, \tan\beta)$ at 90\% (95\%)
C.L., {\it i.e.}, $\chi^2/d.o.f=25.99/18 ~ (28.87/18)$. 
The best-fit point in the {\tt Unconstrained} parameter scan is marked by {\tt\bf u},
and that  in the {\tt Flavor-constrained} parameter scan
by {\tt\bf c}.
Unless other new physics effects relieve the FCNC constraints,
the LHC Higgs signal excludes large $\tb$ region above around 0.6 at 95\% C.L.,
which sets lower bound on the charged Higgs boson mass as $M_{H^\pm} \gsim 800\gev$.

{\renewcommand{\tabcolsep}{10pt}
\begin{table}[t!]
\caption{\label{tab:Type2S1:c:R}The effective couplings and the Higgs signal rates $R$'s
of the best-fit points in the Type II-1 model.}
%\begin{ruledtabular}
\begin{tabular}{|l|l|}
\hline
Type II-1: {\tt Unconstrained} & Type II-1: {\tt Flavor-constrained}\\
 \hline
 $(\chi^2_{\min},~\alpha,~\tb)=(16.16,\, 1.21,\, 0.36)$ & $(\chi^2_{\min},~\alpha,~\tb)=(20.31,\, 0.96,\, 0.50)$ \\ 
\hline \hline
$c_V=-0.76,\quad c_g = 1.12,\quad c_\gm =0.88$ & 
$c_V=-0.47,\quad c_g = 1.35,\quad c_\gm =0.69$\\
$c_b=c_\tau=-0.99,\quad c_t=c_c=1.05$ & $c_b=c_\tau=-0.91,\quad c_t=c_c=1.28$\\
\hline
$R^\ggf_\rr =1.61,\quad R^\vbf_\rr=0.99$ & 
$R^\ggf_\rr =1.67 ,\quad R^\vbf_\rr= 0.61$ \\
$R^\ggf_{WW} = 0.79,\quad R^\vbf_{WW}=0.49,\quad R^\ggf_{ZZ}=0.79$ & 
$R^\ggf_{WW} = 0.51,\quad R^\vbf_{WW}=0.18,\quad R^\ggf_{ZZ}=0.51$ \\
$R^{Vh}_{\bb}=0.63,\quad R^\ggf_\ttau=1.34,\quad R^\vbf_\ttau=0.82$ &
$R^{Vh}_{\bb}= 0.23,\quad R^\ggf_\ttau=1.90,\quad R^\vbf_\ttau= 0.70$\\
 \hline
 \end{tabular}
\end{table}
}

In order to study the physical characteristics of $h^0$ 
at the best-fit point in Type II-1 model,
we show the effective couplings and the Higgs signal rates in Table \ref{tab:Type2S1:c:R}.
We present both best-fit points obtained in the {\tt Unconstrained} and {\tt Flavor-constrained}
parameter space.
For both cases, the $h^0$-$V$-$V$ effective coupling is smaller than the SM value with opposite sign.
In particular, the value of $c_V$ in the flavor-constrained best-fit  point is 
about half of the SM value.
In addition, the top Yukawa coupling is almost the same as the SM value with the same sign.
This combination leads to the effective coupling with a photon pair smaller than the SM value.
Instead $c_g$ increases by about 10\% at the {\tt Unconstrained} best-fit
and about 30\% at the  {\tt Flavor-constrained} best-fit point.
The Higgs signal rates are quite different from the SM values.
Diphoton rate is sizably enhanced in the gluon fusion production,
while reduced for the VBF production.
In particular, the {\tt Flavor-constrained} best-fit point has only 60\% rate for the VBF diphoton channel.
This is attributed to small $c_V$.
With more data at the LHC, this channel will be a major criteria for the Type II-1 model.

\subsection{Scenario-2}
In Scenario-2, the light $h^0$ has not been observed yet
and the new boson is the heavy CP-even $H^0$.
We assume that $A^0$ is heavy and almost degenerate
with $H^\pm$, which suppresses new contributions
to EWPD.
Then we have
\bea
\label{eq:H:c}
c_V &=& \cos (\beta-\alpha),
\quad
c_b =\yh_d^{H},\quad
c_\tau = \yh_\ell^{H},
\quad 
c_t =c_c = \yh_u^{H},
\eea
where $\yh_d^{H}$, $\yh_\ell^{H}$,  and $\yh_u^{H}$
 are in Table \ref{tab:Yukawa}.

The question arises as to why we have not seen the light Higgs boson,
especially at the LEP.
As an $e^+ e^-$ collider, the LEP searched for the Higgs boson
through $e^+ e^- \to Z^* \to Z h \to \llp + jj$.
Despite the tantalizing hint of the Higgs boson with mass around 114.4 GeV
observed by the ALEPH collaborations,
the LEP did not see significant excess over the SM backgrounds~\cite{LEP,LEP2}.
The upper bound on the event rate $|\xi|^2$ was set.
One of
the strongest bounds on $|\xi|^2$ is from flavor-independent jet decay of the Higgs boson.
If the Higgs boson decays with the SM Higgs branching ratios,
$|\xi|^2$ is just the square of the ratio of the $h$-$Z$-$Z$ coupling to the SM value.
In 2HDM, however,
the Higgs boson couplings with fermions also change.
We interprete $|\xi|^2$ as
\bea
|\xi|^2 
=
|c_V|^2 \cdot
\frac{ \br(h^0 \to jj) }{\br(\hsm \to jj)} .
\eea
This LEP constraint, occurring at tree level, is more important than
the flavor constraints at loop level.

{\renewcommand{\tabcolsep}{25pt}
\begin{table}
\caption{\label{tab:Scenario2}Best-fit points for Scenario-2
in Type I, II, X, and Y 2HDM.
For the {\tt Unconstrained} parameter space, we take $-\pi/2<\alpha<\pi/2$
and $\tb>0.1$.
For the {\tt Flavor-constrained}, we only allow $\tb> 0.5$ for Type-II and Y,
$\tb>1.0$ for Type-I and  X.}
\begin{tabular}{|l|l|l|}
\hline
 &{\tt Unconstrained} &{\tt Flavor-LEP-constrained}\\
 \hline
2HDM Type & $(\chi^2_{\min},\,\alpha,\,\tb)$ & $(\chi^2_{\min},\,\alpha,\,\tb)$ \\
\hline\hline
Type I-2 & $(16.11,\,-0.15,\,0.17)$ & $(30.08,\, 0.59,\, 1.01)$ \\ \hline
Type II-2  & $(15.92,\,-0.29,\,0.30)$ &  $(30.87,\, 1.55,\,48.5)$ \\ \hline
Type X-2 & $(15.16,\,-0.35,\,0.27)$ &  $(28.55,\, 1.57 ,\, 49.82)$ \\  \hline
Type Y-2 & $(15.77,\,-0.15,\,0.17)$ & $( 31.91,\, -1.55 ,\, 48.12)$ \\  \hline
\end{tabular}
\end{table}
}

In Table \ref{tab:Scenario2}, we present the best-fit points for Scenario-2.
We scan the parameter space without other constraints ({\tt Unconstrained}),
and with FCNC and LEP bounds ({\tt Flavor-LEP-constrained}).
For the {\tt Unconstrained} best-fit point,
all four types of 2HDM have smaller $\chi^2_{\min}$ than the SM:
small $\tb$ is preferred;
the value of $\alpha$ is negative and small, unlike Scenario-1.

%%%%%%%%%%%%%%%%%%%%%%%%%%%%%%%%%%%%%%%%%%%%%%%%%%%%%%%%%%%%
\begin{figure}[h]
\begin{center}
\includegraphics[width=0.49\linewidth]{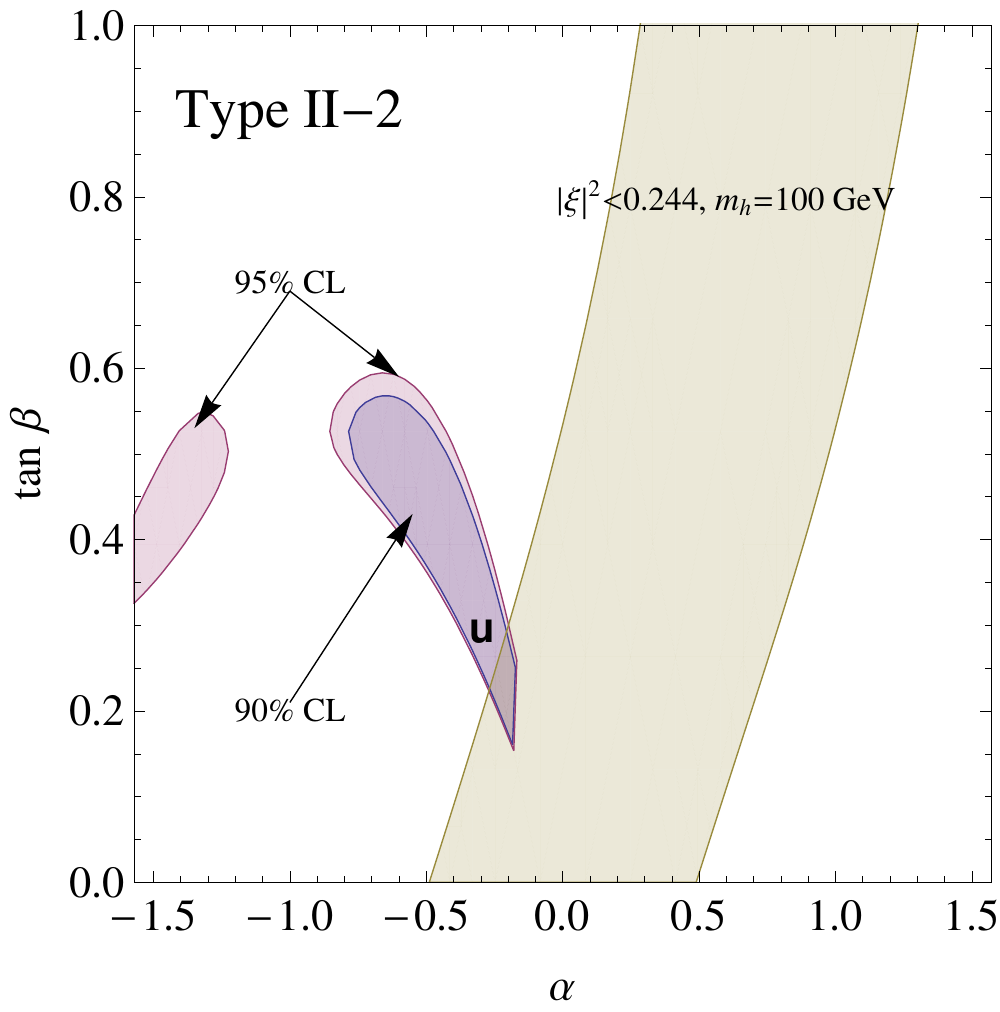}
\end{center}
\caption{Contours for 90 (95)$\%$ in Type II-2 model 
where $h=H^0$.
The best-fit occurs at the point marked by {\tt\bf u} in the {\tt Unconstrained} parameter scan.
The band with the color of light beige is the allowed region by the LEP Higgs search.}
\label{fig:t2s1contour}
\end{figure}
%%%%%%%%%%%%%%%%%%%%%%%%%%%%%%%%%%%%%%%%%%%%%%%%%%%%%%%%%%%

If we impose the FCNC and LEP bounds,
all four models are excluded at 95\% C.L.
The $\chi^2_{\min}$ values are around 30.
Type II-2\!, Type X-2\!, and Type Y-2 models prefer large $\tb$ while Type I-2 model
prefers $\tb\sim 1$.
In this scenario, the LEP bound plays the crucial role.
In Fig.~\ref{fig:t2s1contour},
we present the contours for 90 (95)\% C.L. in Type II-2 model.
There exists a sizable portion of parameter space around negative $\alpha$ and $\tb\sim 0.2-0.6$.
However, the LEP-allowed parameter space is rather away,
yielding only very limited overlap with the Higgs allowed space.
Here we assume that the light $h^0$ mass is 100 GeV,
which corresponds to $|\xi|^2 < 0.244$~\cite{LEP}.
If we lower $m_{h^0}$ further,
the upper bound on $|\xi|^2$ is decreased, leading to stronger LEP bounds.
Even if we increase $m_{h^0}$,
the upper bound on $|\xi|^2$ does not increase, but keeps
almost flat until $m_{h^0}=110\gev$.
For $m_{h^0}\geq 110\gev$,
the ATLAS and CMS data in the diphoton channel
exclude the scenario.
In summary, the condition of $|\xi|^2 < 0.244$ is maximally allowed by the LEP Higgs search.
Type I-3, II-2, and Y-3 models are excluded at 95\% C.L.,
while Type X-3 model is still possibly allowed marginally.

\subsection{Scenario-3}

As motivated by the electroweak precision data,
we consider an exotic scenario where the light CP-even $h^0$ and
the pseudoscalar $A^0$ have almost degenerate mass around 125 GeV.
The observed signal is from two resonances of $h^0$ and $A^0$

{\renewcommand{\tabcolsep}{20pt}
\begin{table}
\caption{\label{tab:Scenario3}Best-fit points for Scenario-3 
in Type I, II, X, and Y 2HDM.
For the {\tt Unconstrained} parameter space, we take $-\pi/2<\alpha<\pi/2$
and $\tb>0.1$.
For the {\tt Flavor-constrained}, we only allow $\tb> 0.5$ for Type-II and Y,
$\tb>1.0$ for Type-I and  X.}
\begin{tabular}{|l|l|l|}
\hline
 & {\tt Unconstrained}& {\tt Flavor-EW-constrained}\\
 \hline
2HDM Type & $(\chi^2_{\min},\,\alpha,\,\tb)$ & $(\chi^2_{\min},\,\alpha,\,\tb)$\\
\hline\hline
Type I-3 &  $(27.52,\, -0.98,\, 1.37)$ & $(29.38,\, -0.68,\,1.62)$ \\ \hline
Type II-3 &  $(28.62,\, 0.23, \, 0.74)$& $(31,03,\, -0.14,\,5.93)$ \\ \hline
Type X-3 & $(15.92,\, -0.34,\, 0.58)$ & $(27.94,\, -0.007,\, 8.32)$ \\ \hline
Type Y-3 &  \multicolumn{2}{c|}{$(30.63,\, -0.75,\, 1.19)$} \\ \hline
%Cell 1 & Cell 2\\
\end{tabular}
\end{table}
}

In this scenario, the $\Delta\rho$ constraint as well as the FCNC ones
is very crucial.
In Table \ref{tab:Scenario3},
we present the best-fit points in Scenario-3.
When scanning the whole parameter space,
the best-fit points show some diversity.
Only the Type X-3 model has $\chi^2_{\rm min}$ much smaller than the SM
while the other three models are already excluded solely 
at 95\% C.L.
If the $\Delta\rho$ constraint applies,
the Type X-3 model 
becomes marginally allowed at 95\% C.L.
And this model is worse than the SM in explaining  the Higgs signals.

%
%%%%%%%%%%%%%%%%%%%%%%%%%%%%%%%%%%%%%%%%%%%%%%%%%%%%%%%%%%%%
%%%%%%%%%%%%%%%%%%%%%%%%%%%%%%%%%%%%%%%%%%%%%%%%%%%%%%%%%%%%
%
\section{Conclusions}
\label{sec:conclusions}
The historic discovery of a new scalar boson of mass 125 GeV
raises a very important question as to whether this new particle 
is the SM Higgs boson.
What is given to us is the entangled combinations of
the productions and decays of the Higgs boson.
The observed new boson may not be the Higgs boson in the SM
but another boson in different models with different couplings.
In the framework of 2HDM,
we answer this question based on 18 different Higgs signal rates observed
by the ATLAS and CMS collaborations at $\sqrt{s}=7\tev$ and $8\tev$.
We comprehensively studied four types of 2HDM with three scenarios.
Scenario-1 is common such that the observed scalar is the light CP-even Higgs boson $h^0$.
Scenario-2 and Scenario-3 are rather exotic, spotting $H^0$ and almost
degenerate $h^0$-$A^0$ for the new boson, respectively.

If only the Higgs signals are relevant,
the global $\chi^2$ fit leads to the conclusion
that all four types of 2HDM in Scenario-1 and Scenario-2
provide better fit to the data than the SM.
The Higgs data prefer small $\tb$ around $0.2-0.3$,
and large positive $\alpha$ (small negative $\alpha$) for Scenario-1 (Scenario-2).
Scenario-3 allows only Type X-3 to have smaller $\chi^2_{\rm min}$ than the SM,
also for small $\tb$.
This small $\tb$ inevitably yields large contributions to FCNC through charged Higgs bosons.
If 2HDM is an effective theory only for the Higgs sector, embedded in a larger theory of new physics,
other new physics effects may relax the flavor constraints.

If we consider flavor constraints more seriously, 
the parameter scan is limited.
In addition there are other important phenomenological bounds:
Scenario-2 should evade the Higgs search at the LEP,
and Scenario-3 should suppress the contribution to $\Delta \rho$.
The global $\chi^2$ fit in the phenomenologically constrained parameter space
excludes all scenarios in all four types of 2HDM,
except for Type II-1 and Type Y-1.
Type II-1 model  has smaller $\chi^2_{\rm min}$ value than the SM, not yet significant at this moment.
Type Y-1 is as good as the SM. 
The Higgs boson at the best-fit point in Type II-1 model has small (about half) coupling with 
the SM gauge bosons, and larger coupling with up-type quarks.
We have enhanced $\rr$ and $\ttau$ modes through gluon fusion,
but reduced $\rr$ mode through VBF production. 
The $\bb$, $WW$, $ZZ$ decay modes are also reduced.
Very different couplings and decay modes of the Higgs boson in Type II-1 model will
play the crucial role in discriminating 2HDM from the SM in the future.

\acknowledgments
This work was supported in part by the National Research Foundation of
Korea (NRF) grant funded by the Korea government of the Ministry of Education, Science
and Technology (MEST) (No. 2011-0003287).
K.Y.L. was supported by the Basic Science Research Program through the NRF funded by MEST (2010-0010916).
S.C.P. is supported by Basic Science Research Program through the NRF of Korea
funded by the MEST
(2011-0010294) and (2011-0029758).
The work of SC and JS is supported by WCU program through the KOSEF funded
by the MEST (R31-2008-000-10057-0).

\end{document}